# Interface effects in spin-dependent tunneling

E. Y. Tsymbal, K. D. Belashchenko, J. P. Velev, S. S. Jaswal,

*Department of Physics and Astronomy and Center for Materials Research and Analysis,*
*University of Nebraska, Lincoln, Nebraska 68588, USA*

M. van Schilfgaarde

*Department of Chemical and Materials Engineering,*
*Arizona State University, Tempe, Arizona 85287, USA*

I. I. Oleynik

*Department of Physics, University of South Florida, Tampa, Florida 33620, USA*

D. A. Stewart

*Cornell Nanoscale Science and Technology Facility,*
*Cornell University, Ithaca, New York 14853, USA*

**Abstract**

In the past few years the phenomenon of spin dependent tunneling (SDT) in magnetic tunnel junctions (MTJs) has aroused enormous interest and has developed into a vigorous field of research. The large tunneling magnetoresistance (TMR) observed in MTJs garnered much attention due to possible application in random access memories and magnetic field sensors. This led to a number of fundamental questions regarding the phenomenon of SDT. One such question is the role of interfaces in MTJs and their effect on the spin polarization of the tunneling current and TMR. In this paper we consider different models which suggest that the spin polarization is primarily determined by the electronic and atomic structure of the ferromagnet/insulator interfaces rather than by their bulk properties. First, we consider a simple tight-binding model which demonstrates that the existence of interface states and their contribution to the tunneling current depend on the degree of hybridization between the orbitals on metal and insulator atoms. The decisive role of the interfaces is further supported by studies of spin-dependent tunneling within realistic first-principles models of Co/vacuum/Al, Co/$Al_2O_3$/Co, Fe/MgO/Fe, and Co/$SrTiO_3$/Co MTJs. We find that variations in the atomic potentials and bonding strength near the interfaces have a profound effect resulting in the formation of interface resonant states, which dramatically affect the spin polarization and TMR. The strong sensitivity of the tunneling spin polarization and TMR to the interface atomic and electronic structure dramatically expands the possibilities for engineering optimal MTJ properties for device applications.

## 1. Introduction

Magnetic tunnel junctions (MTJs) have attracted considerable interest due to their potential application in spin-electronic devices such as magnetic sensors and magnetic random access memories (MRAMs). The diversity of physical phenomena that control the operation of these magnetoresistive devices also makes MTJs very attractive from a fundamental physics perspective. This interest has stimulated tremendous activity in the experimental and theoretical investigations of the electronic, magnetic and transport properties of MTJs (for recent reviews on MTJs see Refs.[1,2]).

A magnetic tunnel junction consists of two ferromagnetic metal layers separated by a thin insulating barrier layer. The insulating layer is so thin (a few nm or less) that electrons can tunnel through the barrier if a bias voltage is applied between the two metal electrodes across the insulator. The most important property of a MTJ is that the tunneling current depends on the relative orientation of magnetization of the two ferromagnetic layers, which can be changed by an applied magnetic field. This phenomenon is called tunneling magnetoresistance (TMR). Since the first observation of reproducible, large magnetoresistance at room temperature by Moodera *et al.* [3], there has been an enormous increase in research in this field. Modern MTJs that are based on 3*d*-metal ferromagnets and $Al_2O_3$ barriers can be routinely fabricated with reproducible characteristics and with TMR values up to 70% at room temperature, making them suitable for applications (see, e.g., Refs.[4,5]).

A very recent discovery of large TMR values in Fe/MgO/Fe(001) and similar MTJs by Parkin *et al.* [6] and Yuasa *et al.* [7] has triggered further interest in the phenomenon of TMR and has opened a new avenue for industrial application of MTJs. These junctions have several advantages over alumina-based MTJs. They are fully crystalline and therefore have well-defined interfaces which can be grown in a more controllable way. They have no disorder-induced localized states in the barrier which are detrimental to TMR [8]. Several experimental groups all over the world including industrial laboratories are now pushing forward research on the electronic, magnetic and transport properties of these and similar MTJs.

The phenomenon of TMR is a consequence of spin-dependent tunneling (SDT), which is an imbalance in the electric current carried by up- and down-spin electrons tunneling from a ferromagnet through a tunneling barrier. SDT was discovered by Tedrow and Meservey [9], who used superconducting layers to detect the spin polarization (SP) of the tunneling current originating from various ferromagnetic electrodes across an alumina barrier [10]. These experiments found a positive SP for all ferromagnetic 3*d* metals. This fact was later explained by Stearns [11], who assumed that the most dispersive bands provide essentially all the tunneling current. Based on this argument and using a realistic band structure of Fe and Ni, Stearns was able to explain experimental values (measured at that time) of the SP for these ferromagnets. Despite the success of Stearns' idea, this model did not provide a clear understanding of the origin of the dominance of the "itinerant" electrons in transport properties.

More recent theoretical studies provided a new insight into the phenomenon of SDT. It was stated that the expected spin dependence of the tunneling current can be deduced from the symmetry of the Bloch states in the bulk ferromagnetic electrodes and the complex band structure of the insulator [12,13]. By identifying those bands in the electrodes that are coupled efficiently to the evanescent states decaying most slowly in the barrier one can make conclusions about the SP of the conductance. It was emphasized that for a broad class of insulating materials the states which belong to the identity representation should have minimum decay rates. This representation is comparable to the *s* character suggesting that *s* bands should be able to couple most efficiently across the interface and decay most slowly in the barrier. For Fe, Co, and Ni ferromagnets the majority-spin states at the Fermi energy have more *s* character than the minority-spin states, which tend to have mainly *d* character. Thus, the majority conductance is expected to be greater than the minority conductance resulting in a slower decay with barrier thickness for the former. These symmetry arguments explain nicely large values of TMR predicted for epitaxial Fe/MgO/Fe junctions [14,15]. These conclusions are also expected to be valid for MTJs with an $Al_2O_3$ barrier which is consistent with the experimental observations [10]. They are also consistent with the earlier hypothesis by Stearns [11].

Despite the undoubted importance, the symmetry arguments have their limitations. First, they assume that the barrier is sufficiently thick so that only a small focused region of the surface Brillouin zone contributes to the tunneling current. For realistic MTJs with barrier thickness of about 1 nm this assumption is usually unjustified. Moreover, for amorphous barriers like alumina where the transverse wave vector is not conserved in the process of tunneling, the entire surface Brillouin zone might contribute almost equally to the conductance. Second, despite the presence of certain selection rules for tunneling, there is no general rule preventing the Bloch states composed mostly of the *d* orbitals from tunneling through the barrier states that have no *d* character. Symmetry strictly forbids tunneling only in special geometries for special values of the wave vector. Therefore, symmetry considerations alone applied to bulk materials are not always sufficient to predict the SP. It is critical to take into account the electronic structure of the ferromagnet/barrier interfaces which, as it was shown experimentally, controls the SP [16].

An important mechanism by which the interfaces affect the SP of the conductance is the bonding between the ferromagnetic electrodes and the insulator [17]. The interface bonding determines the effectiveness of transmission across the interface which can be different for electrons of different orbital character (and/or symmetry) carrying an unequal SP. Also the interface bonding is responsible for the appearance of interface states which, as was predicted theoretically [18], affect the conductance dramatically. Experimentally, the effect of bonding at the ferromagnet/insulator interface was proposed to explain the inversion of the SP of electrons tunneling from Co across a $SrTiO_3$ barrier [19]. The bonding mechanism was also put forward to elucidate positive and negative values of TMR depending on the applied voltage in MTJs with $Ta_2O_5$ and $Ta_2O_5/Al_2O_3$ barriers [20]. Theoretically, strong sensitivity of the magnitude of TMR to the *sp-d* mixing at the ferromagnet/alumina interface was predicted in the presence of imperfectly oxidized Al or O ions [21]. It was found that oxygen deposited on the Fe (001) surface reverses the SP of the density of states (DOS) in vacuum due to the strong exchange splitting of the antibonding oxygen states [22]. It was predicted that an atomic layer of iron-oxide at the interface between Fe and MgO layers greatly reduces TMR in Fe/MgO/Fe junctions due to the bonding between Fe and O [23].

In this paper, we review results of our recent studies of spin-dependent tunneling which show that the SP is primarily determined by the electronic and atomic structure of the ferromagnet/insulator interfaces rather than by bulk properties. In Section 2, we consider a simple tight-binding model which demonstrates that electronic potential and orbital hybridization at the interface essentially control the conductance [24]. In Section 3, we discuss SDT from oxidized Co surfaces through vacuum [25]. We demonstrate that one monolayer of oxygen placed on the fcc Co(111) surface creates a spin-filter effect due to the Co-O bonding. This reverses the sign of the SP from negative for the clean Co surface to positive for the oxidized Co surface. In Section 4, we consider SDT in $Co/Al_2O_3/Co$ junctions [26]. We show that there might be two types of interface O atoms: those which saturate Al bonds and those which are adsorbed by Co. The latter bind strongly to Co creating interface states which enhance the tunneling current in the majority-spin channel, thereby controlling the positive SP. In Section 5, we consider the effect of resonant states on SDT in Fe/MgO/Fe tunnel junctions [27]. We demonstrate that these states are detrimental to TMR at small MgO layer thickness, but can be destroyed by placing a thin Ag layer at the Fe/MgO interface. In Section 6, we analyze spin-dependent tunneling in epitaxial $Co/SrTiO_3/Co$ MTJs [28] and show that the complex band structure of $SrTiO_3$ enables efficient tunneling of the minority $d$-electrons from Co, causing the SP of the conductance across the $Co/SrTiO_3$ interface to be negative in agreement with the experiments of de Teresa *et al*.[19] In Section 7, we make our conclusions.

## 2. Effect of interface bonding within a simple tight-binding model

In order to illustrate the decisive role of interfaces in tunneling properties, we consider, first, a one-dimensional (1D) single-band tight-binding model [24]. Fig.1a shows the geometry of the system which represents a 1D tunnel junction with two metal electrodes separated by an insulating barrier layer. The left electrode consists of a semi-infinite atomic chain with all sites having same on-site atomic energy levels $E_0$ and nearest-neighbor hopping integrals $V_0$. The chain is terminated at a site $s$ which is coupled to the interfacial site $i$ with a hopping integral $V_i$. The site $i$ has an on-site atomic energy $E_i$ and may correspond either to the surface atom of the electrode, or to the interfacial barrier site, or to an "adsorbate" at the metal-barrier interface. In each of these situations, the parameters $E_i$ and $V_i$ are determined by charge transfer and bonding at the interface. The insulator consists of atoms having same energy levels $E_b$ and is coupled to the right electrode, as is shown in Fig.1a. The aim of this simple model is to understand the influence of the hopping integral $V_i$ and the on-site potential $E_i$ at the left interface on tunneling conductance.

To simplify the description we assume that the Fermi level $E_F$ lies well below the bottom of the insulator conduction band, so that the hopping integral $V_b$ between the nearest-neighbor sites in the barrier layer is much less than the barrier height, i.e. $V_b \ll E_b - E_F$. In this limit of a high potential barrier the conductance per spin is given by [29]

$$G(E) = \frac{4\pi^2 e^2}{h} V_b^2 N_i(E) e^{-2\kappa a N} N_r(E), \quad (1)$$

where $G(E)$ is the conductance at a given energy $E$ (Fermi energy), $\kappa = \frac{1}{a}\ln\left(\frac{E_b - E}{V_b}\right)$ is the decay constant, $a$ is the lattice parameter, $N$ is the number of atoms in the insulator, and $N_i(E)$ and $N_r(E)$ are the local DOS at the interface sites $i$ and $r$ respectively. In Eq.(1) the only quantity that depends on the parameters $V_i$ and $E_i$, characterizing the left interface, is the local DOS $N_i(E)$. This quantity can be obtained analytically [24]. Using dimensionless variables $\varepsilon = (E - E_0)/V_0$, $\varepsilon_i = (E_i - E_0)/V_0$, and $w = V_i/V_0$, we find that for the reduced interfacial density of states

$$\rho_i(\varepsilon) \equiv V_0 N_i(E) = -\operatorname{Im} g_i(\varepsilon + i0)/\pi, \quad (2)$$

where $g_i(\varepsilon)$ is the Green's function at atom $i$. For $|\varepsilon| < 2$ it is given by

$$g_i(\varepsilon) = \left[\varepsilon - \varepsilon_i - \frac{w^2}{2}\left(\varepsilon - i\sqrt{4-\varepsilon^2}\right)\right]^{-1}. \quad (3)$$

For $|\varepsilon| < 2$ the Green's function (3) can be obtained by analytic continuation via the upper half-plane.

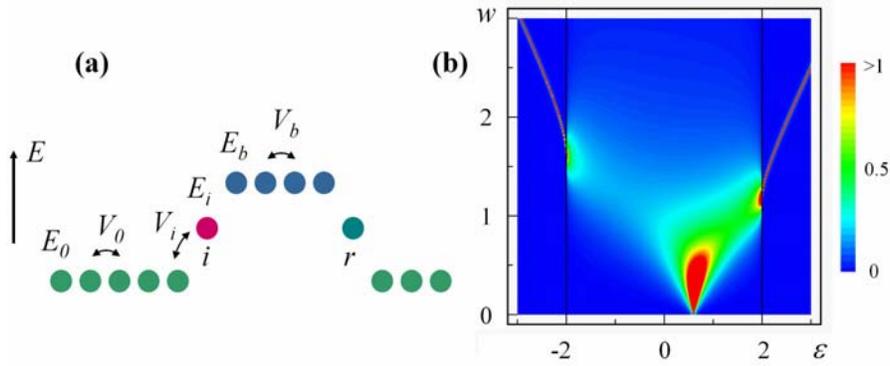

**Fig.1** (a) Geometry of a tunnel junction within a one-dimensional tight-binding model. Vertical positions of atoms reflect the on-site atomic energies. Parameters of the model are described in text. (b) Contour plot of the interface DOS, $\rho_i$, as a function of energy $\varepsilon$ and interface bonding strength $w$ for $\varepsilon_i = 0.6$. The vertical lines at $\varepsilon = \pm 2$ show bulk band edges. After [24].

Fig.1b shows the resulting interface DOS as a function of energy $\varepsilon$ and interface bonding parameter $w$ for an arbitrary choice of $\varepsilon_i = 0.6$. It is seen that the interface

DOS may be strongly enhanced. For strong coupling $w$, this enhancement occurs at band edges. In this case localized states emerge from the continuum. These states correspond to bonding and antibonding orbitals formed by the atom $s$ and its nearest neighbor $i$, modified by the interaction with the bulk band. For weak coupling $w$, the interface DOS is enhanced near the interface atom level $\varepsilon_i = 0.6$. As $w$ decreases starting from large values, the localized level approaches the band edge and then enters the continuum of bulk states, becoming a surface resonance. This is evident from Fig.1b which shows that the interface DOS near this band edge is strongly enhanced. Thus, we see that the magnitude of the interface DOS and consequently the magnitude of the conductance are essentially controlled by the strength of bonding and atomic potential at the interface. The SP of the tunneling current must also be very sensitive to the bonding and potential at the interface due to a different band structure for majority- and minority-spin electrons.

To further illustrate these points, we add the in-plane dispersion to our tight-binding model by considering a simple cubic lattice with nearest-neighbor hopping. For simplicity we assume that all hopping integrals for bonds parallel to the surface are equal to $V_0$. The integrals for perpendicular bonds between the interface $i$ and surface $s$ layers are assumed to be equal to $V_i$, and we again denote $w = V_i / V_0$. In this 3D model, the in-plane component of the wave vector $\mathbf{k}_\parallel$ is conserved and we can use it as a quantum number.

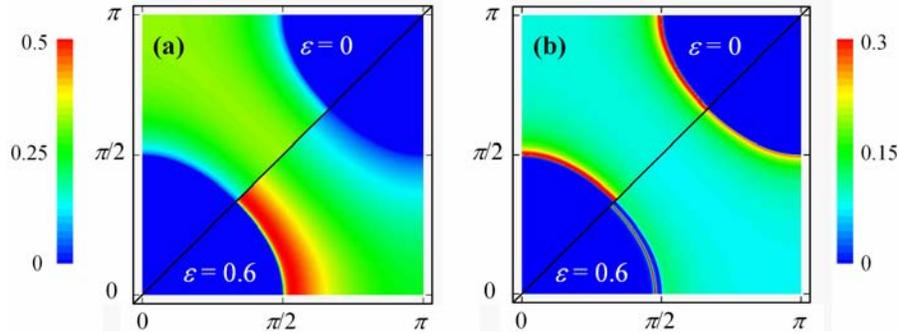

**Fig.2** Contour plots of the interface DOS $\rho_i$ in a quarter of the surface Brillouin zone within the 3D model for $\varepsilon = 0$ and $w = 1$ (a) and $w = 1.5$ (b). The top left corner of each panel shows $N_i$ for $\varepsilon_i = 0$; the bottom right corner, for $\varepsilon_i = 0.6$. After [24].

Fig.2 shows plots of the surface DOS as a function of $\mathbf{k}_\parallel$ for $\varepsilon = 0$ and for four different combinations of $w$ and $\varepsilon_i$. It is clearly seen that the shape of the surface DOS depends strongly on $w$ and $\varepsilon_i$. For $w = 1$ and $\varepsilon_i = 0$, there are no surface perturbations and the DOS varies smoothly with $\mathbf{k}_\parallel$. However, in all other cases the spectral weight is strongly displaced toward one of the edges of the Fermi surface projection (or both edges). For example, for $w = 1$ and $\varepsilon_i = 0.6$, the DOS has a maximum for smaller values of $k_x$ and $k_y$, reflecting the interface resonant state (see Fig.2a, bottom panel).

The above example shows that seemingly small variations in the atomic potentials and hopping integrals near the interface may have a very strong and unexpected effect on the shape of the interface DOS and, hence, on the conductance [30]. Since such variations are common in real materials, the behavior of the interface DOS for bands formed by localized 3$d$ states in transition metals should be very sensitive to the interfacial structure and bonding. As we will see in the next Sections, these effects occur in real MTJs, affecting dramatically the SP of the tunneling current.

## 3. Effect of surface oxidation on spin-dependent tunneling from Co through vacuum

In this section we discuss our results of first-principles calculations of SDT from clean and oxidized Co surfaces through vacuum into Al and demonstrate the crucial role of the bonding between Co and O atoms [25]. We show that a monolayer of oxygen on the Co surface creates a spin-filter effect due to the Co-O bonding by producing an additional tunneling barrier in the minority-spin channel. This reverses the sign of the SP from negative for the clean Co surface to positive for the oxidized Co surface, thus revealing the crucial role of interface bonding in SDT.

The first-principles calculations discussed in this Section, as well as in Sections 4, 5, and 6 are based on a tight-binding linear muffin-tin orbital method (TB-LMTO) in the atomic sphere approximation (ASA) [31] and the local density approximation (LDA) for the exchange-correlation energy. All the atomic potentials are determined self-consistently within the supercell approach. We use a full-potential LMTO (FP-LMTO) method [32] to check the correctness of the ASA in describing the band structure. The principal-layer Green's function technique is applied to calculate the conductance [33,34].

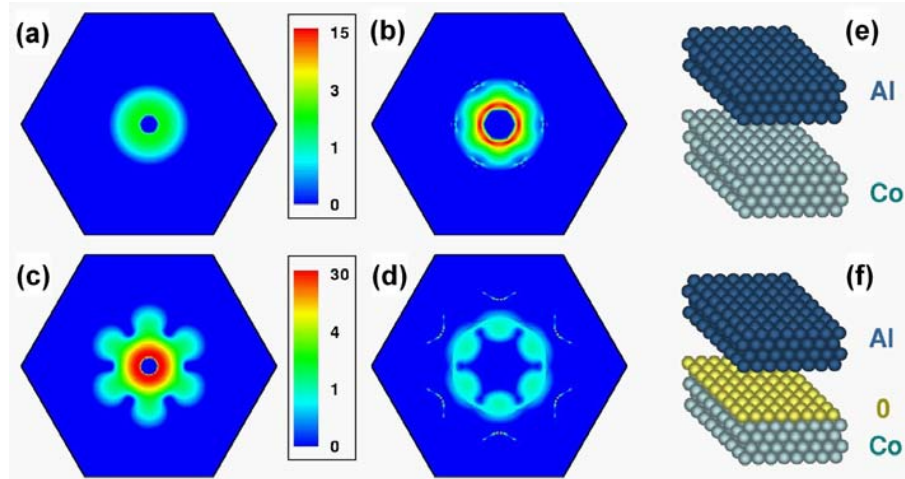

**Fig.3** $\mathbf{k}_\parallel$-resolved conductance (in a logarithmic scale) from clean and oxidized (111) Co surfaces through vacuum into Al: (a) clean surface, majority spin; (b) clean surface, minority spin; (c) oxidized surface, majority spin; (d) oxidized surface, minority spin. The first surface Brillouin zone is shown. Vacuum layer thickness is 2 nm for clean and 1.7 nm for oxidized Co surface. Units are $10^{-11}$ $e^2/h$ for (a), (b) and $10^{-14}$ $e^2/h$ for (c), (d). Geometry of tunnel junctions with clean and oxidized Co surfaces are shown in (e) and (f) respectively. After [25].

We investigate the SP of the current tunneling from a ferromagnetic electrode to a non-magnetic material, Al (111), which serves as a detector of the tunneling SP in the spirit of the Tedrow-Meservey experiments [9,10]. First, we discuss the properties of a Co/vacuum/Al MTJ with a clean Co (111) surface. Figs.3a,b show the $\mathbf{k}_\parallel$-resolved conductance for the majority- and minority-spin electrons within the first surface Brillouin zone of Co (111). The Fermi surface of Co viewed along the [111] direction has holes close to the $\bar{\bar{\Gamma}}$ point with no bulk states in both spin channels, which results in zero conductance in this area. The majority-spin transmission (Fig.3a) varies rather smoothly and is appreciable over a relatively large area of the Brillouin zone. On the other hand, the minority-spin transmission (Fig.3b) has a narrow crown-shaped "hot ring" around the edge of the Fermi surface hole. The analysis of layer and $\mathbf{k}_\parallel$-resolved DOS shows that it is not associated with surface states [18], but rather with an enhancement of bulk $\mathbf{k}_\parallel$-resolved DOS near the Fermi surface edge.

As is seen from Figs.3a,b, the Fermi surface hole is smaller for majority spins. Therefore, the conductance should become fully majority-spin polarized in the limit of very thick barriers. This occurs, however, only at barrier thickness $d \sim 10$ nm, while for typical values of $d \sim 2$ nm the SP is about $-60\%$ and depends weakly on $d$.

An oxidized Co surface is modeled by placing an O monolayer on top of the Co(111) electrode such that O atoms lie in the 3-fold hollow-site positions above the sub-surface Co layer (Fig.3f). To find the equilibrium interlayer distances, we relax the atomic structure of the MTJ using the pseudopotential plane-wave method [35] within the generalized gradient approximation [36] for the exchange-correlation energy. We use the same method in Sections 4 and 5.

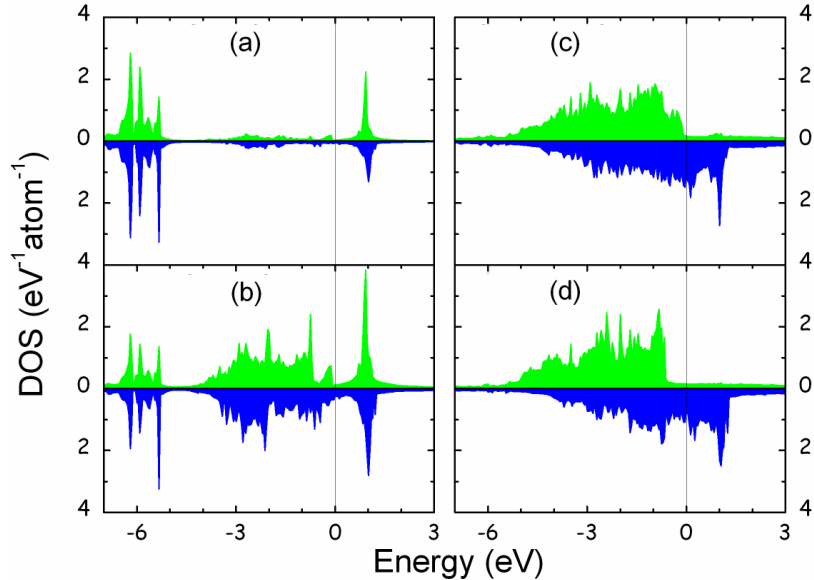

**Fig.4** Layer-resolved density of states as a function of energy for oxidized Co(111) surface: (a) O monolayer; (b) Co monolayer adjacent to O; (c) second Co monolayer; (d) third Co monolayer. Top sub-panels show the majority spin and bottom sub-panels show the minority spin. The Fermi energy lies at zero and is denoted by a solid line.

The oxygen monolayer dramatically changes the electronic structure of the underlying Co monolayer making it almost magnetically-dead. This is evident from Fig.4 which shows the local DOS for the oxidized Co electrode and can be understood as follows. When an O monolayer is deposited onto the Co surface, two sets of bands are formed corresponding to bonding and antibonding mixing of Co and O orbitals. As is seen from Fig.4a,b, the bonding bands lie well below the bulk Co 3$d$ band, whereas the broad band of antibonding states lie around the Fermi energy $E_F$ with a pronounced peak at about 1eV above the $E_F$. As a result of this bonding the local DOS for the monolayer of Co adjacent to O is strongly reduced at $E_F$, so that, according to the Stoner criterion, magnetism in this layer is almost completely suppressed. The magnetic moment of Co atoms in this monolayer is only 0.17 $\mu_B$. As is seen from Figs.4c,d, the DOS of the second and third Co monolayers is very similar to the Co bulk DOS.

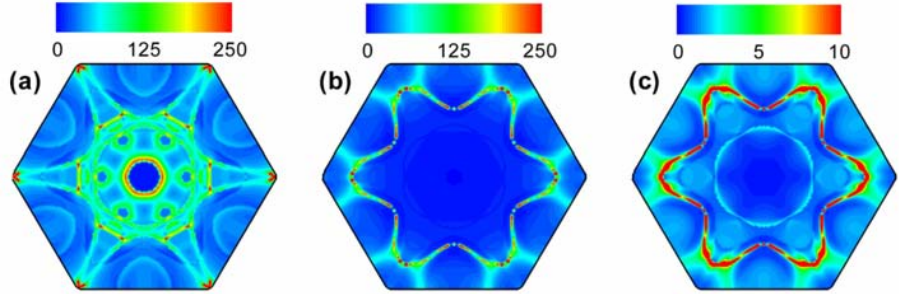

**Fig.5** $\mathbf{k}_\parallel$-resolved minority-spin local DOS at the Fermi energy for the oxidized Co(111) surface (arbitrary units): (a) bulk Co; (b) sub-surface Co monolayer; (c) surface O monolayer. After [25].

Interestingly, the presence of the almost magnetically-dead monolayer of Co at the interface does not kill the spin polarization of the conductance; it rather changes sign of the SP from negative to positive. This fact is evident from Figs.3c,d which shows the $\mathbf{k}_\parallel$-resolved majority- and minority-spin conductance for the oxidized Co surface. The overall reduction in the conductance being the consequence of oxidation is accompanied by the dominant suppression of the conductance of minority-spin electrons.

The origin of this behavior can be understood from Fig.5 which displays the $\mathbf{k}_\parallel$- and layer-resolved minority-spin DOS at the Fermi energy. For bulk Co the Fermi surface edges are strongly emphasized in the DOS (Fig.5a). One of them representing a ring around the $\overline{\Gamma}$ point and corresponding to smaller $\kappa(\mathbf{k}_\parallel)$ dominates the minority-spin conductance for the clean Co surface (compare to Fig.3b). The oxidation results in the strong covalent bonding between Co and O atoms at the surface producing an antibonding band which is clearly seen in the $\mathbf{k}_\parallel$-resolved DOS for Co and O surface monolayers (Figs. 5b,c). This resonant band appears only for minority-spin electrons due to the selection rule that prevents them from mixing with bulk states in this channel (for details, see Ref. [25]). The minority-spin surface states lie rather far from the center of the surface Brillouin zone, and hence are suppressed by the vacuum decay, as is seen from the minority-spin conductance in Fig.3d. At the same time, the interface bonding removes the spectral weight from the center of the Brillouin zone. As a result, the bulk minority-spin states responsible for most tunneling transmission from the clean surface encounter a band gap in the surface Co and O layers which is equivalent to adding an additional tunneling barrier. Thus, the tunneling conductance for the MTJ with the oxidized Co surface is fully dominated by the majority-spin channel, resulting in SP of about +100%.

Experimentally, the reversal of the SP associated with surface oxidation may be detected using spin-polarized STM measurements [37]. Since the ferromagnetic tip is sensitive to the SP of the total local DOS above the surface (see, e.g., Ref. [38]), the TMR in the system surface/vacuum/tip should change sign when the Co surface is oxidized. In other words, for the clean Co (111) surface the tunneling current should be higher when the magnetizations of the tip and the surface are aligned parallel (the dominating minority channel is then open), but for the oxidized surface it should be higher for the antiparallel configuration.

## 4. Positive spin polarization driven by O adsorption in $Co/Al_2O_3/Co$ tunnel junctions

In this section we consider spin-dependent tunneling in $Co/Al_2O_3/Co$ MTJs [26]. Assuming crystalline epitaxy at the interface between Co and $Al_2O_3$ we consider two fully-relaxed atomic configurations of the O-terminated interface that differ only by the presence or absence of an adsorbed oxygen atom at the interface. We show that these structures exhibit opposite signs of the spin polarization of the tunneling current, reflecting features of the electronic structure and bonding at the $Co/Al_2O_3$ interface and, thereby, showing the crucial role of the interface in controlling the spin polarization.

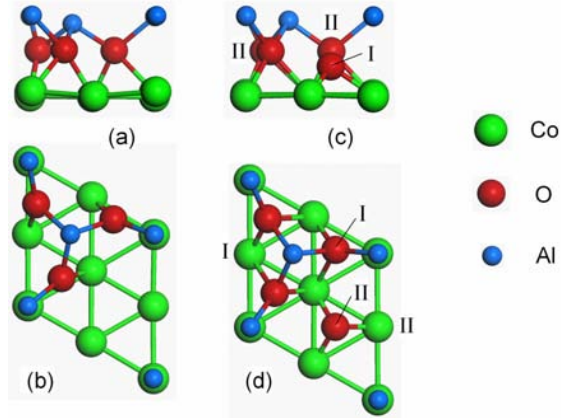

**Fig.6** Interface structure of the $Co/Al_2O_3/Co$ MTJ for model 1 (a, b) and model 2 (c, d). Panels (a) and (c) show "front" views from a direction normal to the 3-fold axis; panels (b) and d) show "top" views along the 3-fold axis. There are two types of Co and O atoms at the interface for model 2: three O(I) atoms, one O(II) atom, one Co(I) atom, and three Co(II) atoms per unit cell. After [26].

The interface structure for the two structural models is shown in Fig.6. Both models have (111)-oriented fcc Co electrodes. Model 1 (Figs.6a,b) represents the O-terminated Co/Al$_2$O$_3$/Co structure studied previously [39]. In this structure the interface contains three oxygen atoms per unit cell. The oxygen atoms are located close to the bridge adsorption sites of the Co surface. These oxygen atoms participate in bonding with the two adjacent Al atoms, making the bonds of the latter fully saturated. Model 2 (Figs.6c,d) adds an additional O atom in the 3-fold hollow site. This O atom and the neighboring Co atoms are labeled "II" in Fig.6c,d, whereas the other O and surface Co atoms are labeled "I". Structural sites occupied by O(I) and O(II) atoms are very dissimilar. O(II) atoms lie much closer to the Co surface compared to O(I) atoms. In fact, within a few hundredths of an angstrom these sites are identical to the O adsorption sites for monolayer coverage. Thus, O(II) atoms are more strongly coupled to Co than to O(I) and Al atoms. Qualitatively, O(II) atoms may be regarded as "Co adsorbates", while O(I) atoms, as "Al$_2$O$_3$-terminating".

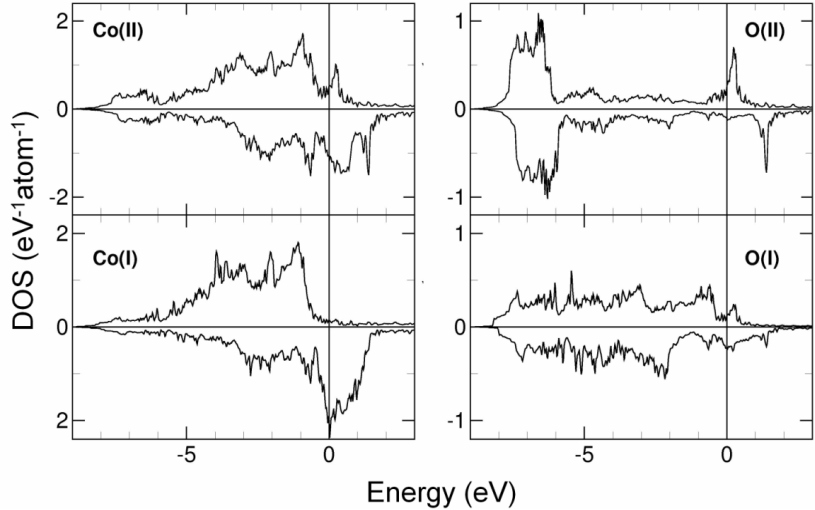

**Fig.7** Local densities of states per atom for interfacial atoms in model 2. In each panel, top half shows the majority-spin DOS, and bottom half, the minority-spin DOS per atom. The vertical line denotes the position of the Fermi level. After [26].

This distinction is evident in the local DOS for the interfacial atoms shown in Fig.7. Similar to the Co (111) surface with an adsorbed oxygen monolayer, Co(II) and O(II) atoms in model 2 form bonding and antibonding orbitals which are clearly seen in the local DOS plots shown in Fig.7. The bonding states lie below the bottom of the Co 3$d$-band, while the antibonding states are slightly above the Fermi level. Some of the DOS weight is removed from the Fermi level to these hybridized states, so that the Stoner criterion for Co(II) atoms is weakened. The magnetic moments at the interface layer are 2.09 $\mu_B$ for Co(I) and 1.30 $\mu_B$ for Co(II). While the magnetic moment of Co(II) atoms is notably reduced, this effect is much smaller compared to the oxidized Co surface, because in model 2 there is only one "adsorbed" O(II) atom per three Co(II) atoms. The local DOS for Co(I) atoms remains quite similar to bulk Co, while the local DOS for O(I) atoms shows a small but notable "echo" of the Co(II)-O(II) antibonding states.

In order to obtain the spin asymmetry of the conductance for these tunnel junctions, we calculate the spin-resolved conductance for the parallel orientation of electrodes. For model 1 we find that the majority-spin conductance $G_{\uparrow\uparrow}$ is 0.0042 $e^2/h$ per unit cell area. It is smaller than the minority-spin conductance $G_{\downarrow\downarrow}$ which is 0.023 $e^2/h$ per cell area. This implies that the spin polarization $P = \frac{G_{\uparrow\uparrow} - G_{\downarrow\downarrow}}{G_{\uparrow\uparrow} + G_{\downarrow\downarrow}}$ is negative and equals –70%. Note that, although this quantity is not directly measurable, it correlates with the measurable spin polarization.

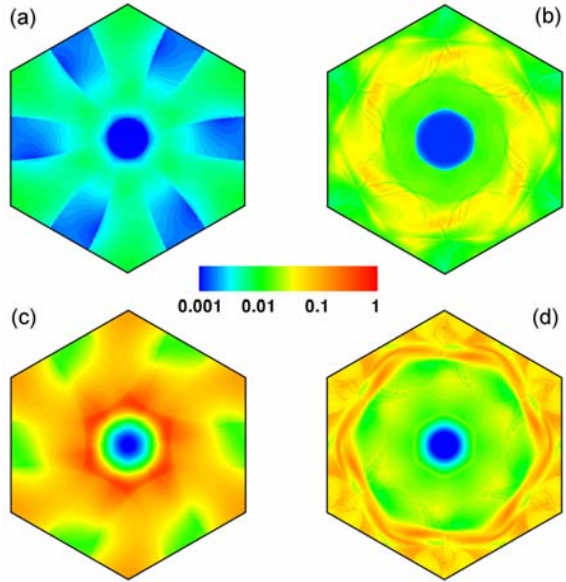

**Fig.8** $k_\parallel$-resolved conductance per unit cell area in a logarithmic scale in units of $e^2/h$ for Co/Al$_2$O$_3$/Co tunnel junctions with different interface structures: (a) model 1, majority spin; (b) model 1, minority spin; (c) model 2, majority spin; (d) model 2, minority spin. After [26].

This situation changes dramatically when an additional O atom is placed at the interface. We find that the model 2 exhibits a reversal of the spin polarization from negative to positive. The total conductances per cell area are $G_{\uparrow\uparrow}$ = 0.087 $e^2/h$ and $G_{\downarrow\downarrow}$ = 0.045 $e^2/h$, and the $P$ = +32%. This is similar to the vacuum barrier case, showing that the deposition of a monolayer of O on the Co (111) reverses the spin polarization compared to the pure Co surface (Section 3). However, the mechanism of the reversal in the case of the Al$_2$O$_3$ barrier is different as explained below.

Fig.8 shows the calculated spin- and $k_\parallel$-resolved conductance within the two models. We see that the minority-spin conductance is qualitatively similar for the two models (Figs.8b,d), whereas the majority-spin conductance is quite different (Figs.8a,c). For model 2 we observe that the majority-spin tunneling current is dominated by a rather narrow hexagonally shaped region encircling the central region of the low conductance (Fig.8c). This feature is not present in model 1 (Fig.8a) and is induced by interface states [26]. It is the consequence of the antibonding Co(II)-O(II) states present at the Fermi energy for majority-spin electrons (see Fig.7). The corresponding minority-spin states lie more than 1 eV above the Fermi level due to exchange splitting, and hence do not contribute to the conductance.

Thus, we see that the interface adsorption of O atoms is responsible for the positive spin polarization of the tunneling current in Co/$Al_2O_3$/Co tunnel junctions. The bonding of the adsorbed O atoms with Co produces antibonding Co-O states that are present at the Fermi level in the majority-spin channel. These interface states moderately mix with the bulk states, forming interface resonances which strongly assist tunneling. On the other hand, the minority-spin antibonding Co-O states lie above the Fermi energy due to exchange splitting and do not affect the conductance. Our results suggest that the common argument of the dominant *s*-electron contribution to tunneling, which is often used to explain the positive spin polarization of the alumina-based tunnel junctions is not fully justified. The interfacial adsorption of oxygen may be the major factor resulting in the positive spin polarization as is observed in experiment.

## 5. Interface resonant states in Fe/MgO/Fe tunnel junctions

In this section we consider epitaxial Fe/MgO/Fe(001) magnetic junctions [27]. Experimental results show that in these junctions TMR decreases precipitously for barrier thickness below 2 nm [7]. Here we demonstrate that the reduction of TMR at small barrier thickness is controlled by the minority-spin interface band. We predict that a monolayer of Ag epitaxially deposited at the interface between Fe and MgO suppresses tunneling through this interface band and may thus be used to enhance TMR for thin barriers.

To calculate the tunneling conductance in Fe/MgO/Fe(001) MTJs we use the atomic structure given in Ref.[14]. Results of the calculation are shown in Figs.9a-c which display the spin-resolved conductance for the MTJ with 4 monolayers (MLs) of MgO for the parallel and antiparallel magnetization. As is evident from Fig.9b, the conductance in the minority-spin channel is strongly enhanced. This enhancement is due to a resonant interface band which is clearly seen in red in this figure. The presence of this interface band can also be seen in the energy-resolved DOS as a narrow peak near the Fermi level (e.g., Fig.3b in Ref.[14]) and in the $\mathbf{k}_\parallel$-resolved local DOS as a curve of finite width in the interface Brillouin zone [27]. The enhancement of the conductance is most pronounced for small barrier thickness, because the interface band lies away from the $\bar{\Gamma}$ point, and therefore the resonant contribution to the transmission decays faster with barrier thickness compared to the non-resonant contribution. We find that for MgO thickness smaller than 6 MLs the contribution from minority-spin electrons in the parallel configuration becomes higher than that from majority-spin electrons.

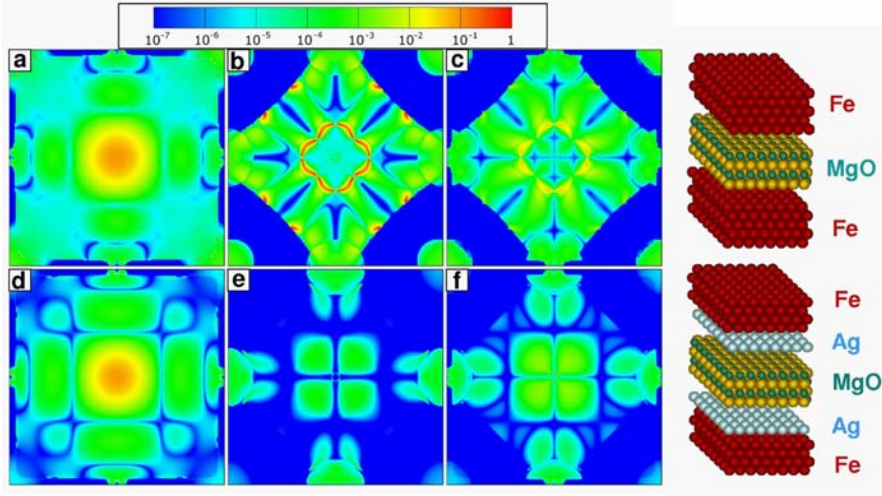

**Fig.9** Conductance (in units of $e^2/h$) as a function of $\mathbf{k}_\parallel$. (a)-(c): Fe/MgO/Fe junction with 4 MgO MLs; (d)-(f): Same junction with Ag interlayers. (a), (d): Majority spins; (b), (e): Minority spins; (c), (f): Each spin channel in the antiparallel configuration. After [27].

An important property of the minority-spin interface resonances is that they strongly contribute to the conductance in the parallel configuration only for ideal, symmetric junctions, and only at zero bias. This follows from strong sensitivity of the location of the interface resonances in the interface Brillouin zone to energy [27]. A bias voltage of the order of 0.01 eV is sufficient to destroy the matching between resonances at the two interfaces. We checked this by calculating the conductance for a small bias voltage and found that at 0.02 eV bias voltage the conductance becomes fully dominated by majority-spin electrons. Disorder also tends to break the matching of the interface resonances even at zero bias. Within the real-space picture, disorder localizes the interface resonances differently at the two interfaces, so that tunneling between them is suppressed. Within the reciprocal-space picture averaging over disorder configurations increases the damping $\gamma$ of the interface resonances. Assuming that $\gamma$ is much greater than the natural damping $\gamma_0$ due to mixing with bulk states, interface resonances are smeared out over an area of the interface Brillouin zone proportional to $\gamma$. If the spectral weight is approximately conserved (which is true if the real part of the self-energy due to disorder is relatively small), the magnitude of the $\mathbf{k}_\parallel$-resolved surface DOS at its maximum decays as $1/\gamma$. For the parallel magnetization with matching interface resonances, the transmission function then decreases as $1/\gamma^2$, and the conductance as $1/\gamma$. Therefore, disorder reduces the conductance in this channel by a factor of $\gamma/\gamma_0$. Thus, we argue that in real Fe/MgO/Fe MTJs the minority-spin channel in the parallel configuration is closed.

Unlike the parallel configuration, the interface resonances *do* contribute to the conductance in the antiparallel configuration, where they tunnel into majority-spin states of the other electrode. The latter have no fine structure in the Brillouin zone, and hence the conductance is weakly sensitive to a potential mismatch at the two interfaces which might occur in real junctions. The enhanced contribution of these interface resonances, which is clearly seen in Fig.9c, leads to the decrease of TMR at low barrier thickness.

In order to enhance TMR for thin MgO barriers we propose to use thin epitaxial Ag interlayers deposited at the Fe/MgO interfaces. Since the lattice parameter of Ag is close to both Fe and MgO lattice parameters, epitaxial Fe/Ag/MgO/Ag/Fe(001) tunnel junctions are feasible. It is known that an epitaxial Ag overlayer on Fe(001) surface notably modifies the electronic structure of the surface states [40]. If the minority-spin interface DOS is reduced by Ag, the antiparallel conductance will be suppressed. On the other hand, the majority-spin conductance should not be strongly affected due to almost perfect transmission through the Fe/Ag(001) interface [41]. This is the rationale for using Ag interlayers.

Figs.9d-f show the $k_\parallel$- and spin-resolved conductance of Fe/MgO/Fe junctions with Ag interlayers. Not unexpectedly, the majority-spin conductance is weakly affected by the Ag interlayers, whereas the minority-spin conductance and the spin conductance in the antiparallel configuration change dramatically. The most pronounced difference for the latter two is the disappearance of the interface resonances that dominated the conductance of the Fe/MgO/Fe junction that lacked Ag interlayers (compare Figs.9b,c and Figs.9e,f). This strong change occurs due to the Fe-Ag hybridization which makes the interface resonant band more dispersive and removes the Fermi level crossing responsible for the highly conductive resonant states. The significant reduction of the conductance in the antiparallel configuration leads to a dramatic enhancement of the TMR. Thus, Ag interlayers practically eliminate the contribution from the interface resonances and enhance TMR for thin barriers.

Thus, interface resonant states in Fe/MgO/Fe(001) tunnel junctions contribute to the conductance in the antiparallel configuration and are responsible for the decrease of TMR at small barrier thickness, which explains the experimental results of Yuasa *et al.*[7] Depositing thin Ag interlayers at the Fe/MgO interfaces suppresses tunneling through these resonant states and thereby enhances the TMR for thin barriers. These results further support our main statement about the decisive role of the electronic structure at the interface in controlling the spin polarization and TMR.

## 6. Negative spin polarization in Co/SrTiO$_3$/Co tunnel junctions

In this section we analyze spin-dependent tunneling in epitaxial Co/SrTiO$_3$/Co(001) MTJs [28]. The motivation for this study is the work of de Teresa *et al*. [19], who found that the tunneling SP depends on the insulating barrier. They used a half-metallic La$_{0.7}$Sr$_{0.3}$MnO$_3$ (LSMO) as a spin detector in Co/Al$_2$O$_3$/LSMO and Co/SrTiO$_3$/LSMO MTJs. Since LSMO has only majority states at the Fermi energy, its tunneling SP is positive and close to 100%, regardless of the insulating barrier. As expected, Co/Al$_2$O$_3$/LSMO MTJs was found to have a normal TMR. Surprisingly, Co/SrTiO$_3$/LSMO MTJs showed an *inverse* TMR. De Teresa *et al*.

proposed that the SP of the Co/SrTiO$_3$ interface must be negative, opposite to that of the Co/Al$_2$O$_3$ interface. In this section we demonstrate that the complex band structure of SrTiO$_3$ enables efficient tunneling of the minority *d*-electrons from Co, causing the SP of the conductance across the Co/STO interface to be negative which explains experiments of de Teresa *et al*.

Our method employs the structural model of an epitaxial Co/SrTiO$_3$/Co(001) MTJ obtained by Oleinik *et al*. [42] and shown in Fig.10a. The lattice parameters of bulk fcc Co and bulk SrTiO$_3$ in its equilibrium perovskite structure have a 10% lattice mismatch which would normally prevent epitaxial growth. However, good metals usually accommodate various lattice structures with only a small energetic penalty because their binding energy depends primarily on density. Therefore, in the calculation of Oleinik *et al.* the strained Co layer proceeded along the Bain path from the fcc to the slightly distorted bcc structure. It is therefore likely that the top Co electrode in the experiments of de Teresa *et al*. [19] grew in the bcc phase on SrTiO$_3$, and that fully crystalline bcc Co/SrTiO$_3$/Co(001) MTJs may be grown on a suitable substrate (see, e.g., experimental papers [43,44] on growth of bcc Co).

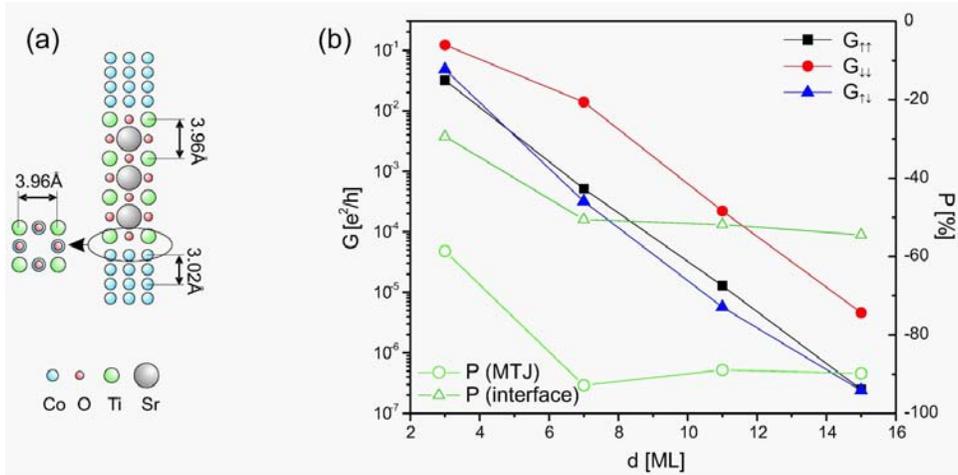

**Fig.10** (a) Schematic graph of the most stable interface structure of the Co/SrTiO$_3$/Co(001) MTJ taken from Ref.[42]. Projections on two perpendicular planes are shown. (b) Conductance *G* (solid symbols) and spin polarization *P* (open symbols) versus barrier thickness *d* for Co/SrTiO$_3$/Co magnetic tunnel junctions. *P*(MTJ) denotes the spin polarization obtained by calculating the majority- and minority-spin conductance for parallel-aligned MTJ. *P*(interface) denotes the spin polarization obtained for the Co/SrTiO$_3$ interface as described in text. After [28].

The spin-resolved conductance of a Co/SrTiO$_3$/Co MTJ is shown in Fig.10b as a function of barrier thickness. It is seen that the conductance decreases exponentially with a similar decay length for parallel and antiparallel configuration of the electrodes. The conductance of the minority-spin channel in the parallel configuration $G_{\downarrow\downarrow}$ is greater than that of the majority-spin channel $G_{\uparrow\uparrow}$, or of any spin channel in the antiparallel configuration $G_{\uparrow\downarrow}$. The SP of the conductance in the parallel configuration, $P = \dfrac{G_{\uparrow\uparrow} - G_{\downarrow\downarrow}}{G_{\uparrow\uparrow} + G_{\downarrow\downarrow}}$, is negative for all barrier thicknesses (see P(MTJ) in Fig.10b). Moreover, except for the thinnest barrier of 3 monolayers (ML), the *P* is almost constant at –90%, and the TMR is very large (about 2000% for 7 and 11 MLs, and about 1000% for 15 MLs of $SrTiO_3$).

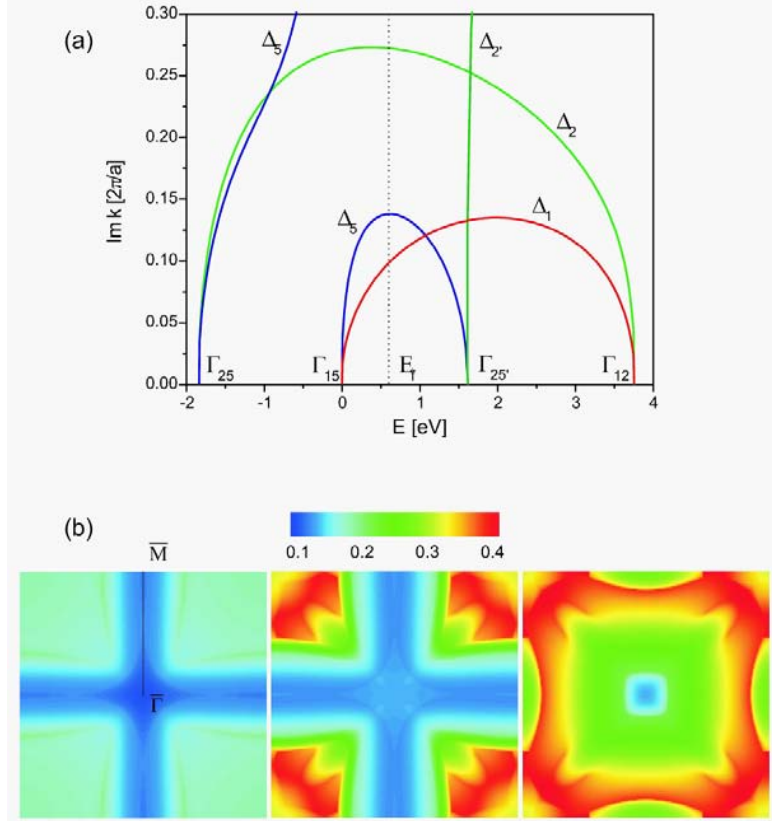

**Fig.11** (a) Complex band structure of $SrTiO_3$ at the $\overline{\Gamma}$ point. The position of the Fermi level $E_f$ in a Co/$SrTiO_3$/Co MTJ is shown by a dashed line. (b) Three lowest decay rates (in units of $2\pi/a$) of the evanescent states in $SrTiO_3$ as a function of $\mathbf{k}_\parallel$ at the Fermi energy. After [28].

The fact that the tunneling current is dominated by minority-spin electrons can be explained by taking into account the band structure of bcc Co and decay rates of the Co states in $SrTiO_3$. The majority-spin 3*d* band in bcc Co is filled, so that the DOS at the Fermi level has a large negative SP [45]. If the 3*d* states could efficiently tunnel through the barrier, the tunneling SP would also be negative. As seen from the complex band structure of $SrTiO_3$, shown in Fig.11a at the $\overline{\Gamma}$ point ($\mathbf{k}_\parallel$=0), the $\Delta_5$ and $\Delta_1$ states have comparable decay rates in the gap of $SrTiO_3$. Therefore, both the majority-spin $\Delta_1$ band and the minority-spin $\Delta_5$ band of bcc Co [45] can tunnel efficiently through the $SrTiO_3$ barrier.

While the $\bar{\Gamma}$ point analysis is instructive, it is not sufficient because the conductance is not dominated by this point. This fact can be understood from Fig.11b, showing the three lowest decay rates of the evanescent states at the Fermi energy. It is seen that a very large area of the Brillouin zone, forming a cross pattern along the $\bar{\Gamma}$–M directions, exhibits two lowest decay rates that are very close to those at the $\bar{\Gamma}$ point. Clearly, at large barrier thickness the states lying in this "cross" area should dominate the conductance. This feature is in sharp contrast to *sp*-bonded insulators like MgO and $Al_2O_3$ where the decay rate has a deep parabolic minimum in the vicinity of the $\bar{\Gamma}$ point. This difference is due to the conduction band of $SrTiO_3$ which is formed by fairly localized 3*d* states of Ti instead of free-electron-like states of a metal atom in simple oxides. Therefore, the minority-spin *d* states which have much larger DOS at the Fermi energy than the majority-spin states dominate the conductance providing a negative SP of the tunneling current in $Co/SrTiO_3/Co$ MTJs.

The analysis of $\mathbf{k}_\parallel$-resolved conductance over the interface Brillouin zone for the parallel alignment of $Co/SrTiO_3/Co$ MTJs reveals a significant mismatch between the majority and minority spin channels [28]. Since tunneling electrons must traverse both interfaces and $\mathbf{k}_\parallel$ is conserved, this makes the conductance in the antiparallel configuration much smaller than the conductance in the parallel configuration resulting in a very large TMR.

Now we make a quantitative comparison of our results with the experiments of de Teresa *et al*. [19] who found that the SP of the $Co/SrTiO_3$ interface is –25%. We determine the SP of the interface from the metal-induced DOS in the barrier. In doing this, we approximate the LMSO electrode as an ideal spin analyzer, similar to the Tersoff-Hamann model for an STM tip [46], and assume that the DOS in the barrier is simply the sum of DOS induced by the left and right electrodes (this is valid as long as the barrier is not too thin). Since in our case pure surface states are absent, the entire barrier DOS is metal-induced. Therefore, we can use the total DOS in the middle of the barrier with no ambiguity. The SP of the $Co/SrTiO_3$ interface obtained in such a way is close to –50% and is almost independent of barrier thickness (see *P*(interface) in Fig.10b). Thus, our model explains the negative value of the SP of the $Co/SrTiO_3$ interface obtained by de Teresa *et al*.; some quantitative difference may be related to the effects of disorder unavoidable in experiment.

Thus, the large negative tunneling spin polarization of the $Co/SrTiO_3$ (001) interface is due to the complex band structure of $SrTiO_3$ which is formed from localized 3*d* states of Ti and hence allows efficient tunneling of the minority *d* electrons of Co. This behavior is a significant departure from the mechanism of tunneling in MTJs based on *sp*-bonded insulators supporting conduction of majority-spin electrons.

### 7. Conclusions

This paper emphasizes the critical role of electronic and atomic structure of interfaces in spin-dependent tunneling in magnetic tunnel junctions. A simple single-band tight-binding model shows that small variations in atomic potentials and bonding near the interface have a very strong effect on the interface DOS and on the conductance. Such variations are common in real materials and the behavior of the interface DOS for bands formed by localized 3$d$ states in transition metals is very sensitive to the interfacial structure and bonding. We have, indeed, found such effects in realistic first-principles models for Co/vacuum/Al, Co/Al$_2$O$_3$/Co, Fe/MgO/Fe, and Co/SrTiO$_3$/Co MTJs, affecting dramatically the SP of the tunneling current and TMR. In particular, for Co/vacuum/Al junctions we found that depositing a monolayer of oxygen on the Co (111) surface reverses the spin polarization from –60% to almost +100% due to the formation of surface bands that mix well with majority-spin Bloch states but create an additional tunneling barrier for minority-spin Bloch states. For Co/Al$_2$O$_3$/Co MTJs, we demonstrated that a somewhat similar effect is produced by interfacial adsorption of oxygen at the Co/Al$_2$O$_3$ interface. Contrary to the Co/vacuum/Al MTJ, however, the spin dependence in this case is related to the exchange splitting of the antibonding Co-O states. Our results for Co/Al$_2$O$_3$/Co MTJs suggest a possible explanation of the experimentally observed positive spin polarization in these junctions. For Fe/MgO/Fe(001) MTJs we predicted that for small barrier thickness the minority-spin resonant bands at the two interfaces make a significant contribution to the tunneling conductance for the antiparallel magnetization which explains the experimentally observed decrease in TMR for thin MgO barriers. A monolayer of Ag epitaxially deposited at the interface between Fe and MgO suppresses tunneling through the interface band and may thus be used to enhance the TMR. For Co/SrTiO$_3$/Co MTJs with bcc Co(001) electrodes we predicted a very large TMR, originating from a mismatch of majority- and minority-spin states contributing to the conductance. In agreement with experimental data we found that the spin polarization of the tunneling current across the Co/SrTiO$_3$ interface is negative. We attributed this property to the complex band structure of SrTiO$_3$ which is formed from localized 3$d$ states of Ti and hence allows efficient tunneling of the minority $d$ electrons of Co.

The strong sensitivity of the tunneling spin polarization and tunneling magnetoresistance to the interface atomic and electronic structure makes the quantitative description of transport characteristics of MTJs much more complicated; however, it broadens dramatically the possibilities for altering the properties of MTJs. In particular, by modifying the electronic properties of the ferromagnet/insulator interfaces it is possible to engineer MTJs with properties desirable for device applications.

**Acknowledgements**

This work was supported by NSF (DMR-0203359 and MRSEC DMR-0213808) and the Nebraska Research Initiative. IIO was supported by NSF (CCF-0432121). The calculations were performed using the Research Computing Facility of the University of Nebraska-Lincoln.